\documentclass[useAMS,usenatbib]{mn2e}

\usepackage{xspace}
\usepackage[]{natbib}
\usepackage{times}
\usepackage{hyperref} 
\usepackage{amsmath} 
\usepackage{amssymb} 
\usepackage{amsfonts} 
\usepackage{graphicx} 
\usepackage{indentfirst} 
\usepackage{pdflscape} 
\usepackage{placeins} 
\usepackage{nicefrac} 
\usepackage{afterpage} 
\usepackage{rotating} 
\usepackage{booktabs} 
\usepackage{rotating} 
\usepackage{color} 

\newcommand{\R}{\,{\mathbb R}}

\newcommand{\dx}{\,{\mathrm d}x}

\begin{document}

\title{Umbrella sampling: a powerful method to sample tails of distributions}

\author[Matthews et al.]{Charles Matthews$^{1}$\thanks{E-mail: c.matthews@uchicago.edu}, Jonathan Weare$^1$, Andrey Kravtsov$^{2,3}$, Elise Jennings$^{4,5}$\\
\\
$^{1}$Department of Statistics, and James Frank Institute, The University of Chicago, Chicago IL 60637 \\
$^{2}$Department of Astronomy \& Astrophysics, and Kavli Institute for Cosmological Physics, University of Chicago, Chicago IL 60637 \\
$^{3}$Enrico Fermi Institute, The University of Chicago, Chicago, IL 60637, USA \\
$^{4}$Theoretical Astrophysics Division, Fermi National Laboratory, Batavia, IL, USA \\
$^{5}$ Leadership Computing Facility,  Argonne National Laboratory, Argonne, IL, 60439, USA}
\pagerange{\pageref{firstpage}--\pageref{lastpage}} \pubyear{2016}

\maketitle

\label{firstpage}

\begin{abstract}
We present the umbrella sampling (US) technique and show that it can be used to sample extremely low probability areas of the posterior distribution  that may be required in statistical analyses of data. In this approach 
sampling of the target likelihood is split into sampling of multiple biased likelihoods confined within individual umbrella windows. 
We show that the US algorithm is efficient and highly parallel and that it can be easily used 
with other existing MCMC samplers. The method allows the user to capitalize on their intuition and define umbrella windows and increase sampling accuracy along specific directions in the parameter space. Alternatively,  one can define umbrella windows using an approach similar to parallel tempering. We provide a public code that implements umbrella sampling as a standalone python package.
We present  a number of tests illustrating the power of the US method in sampling low probability areas of the posterior and show  that this ability allows a considerably more robust sampling of multi-modal distributions compared to the standard sampling methods. We also present  an application of the method in a real world 
example of deriving cosmological constraints using the supernova type Ia data. We show  that  umbrella sampling can sample the posterior accurately down to the $\approx 15\sigma$ credible region in the $\Omega_{\rm m}-\Omega_\Lambda$ plane, while for the same computational work the affine-invariant MCMC sampling implemented in the {\tt emcee} code samples the posterior reliably only to $\approx 3\sigma$. 
\end{abstract}


\section{Introduction}

Markov Chain Monte Carlo (MCMC), based on the Metropolis-Hastings class of algorithms \citep{metropolis_etal53,hastings70}, has enjoyed  great success in a wide range of fields from astrophysics and physics \citep[see, e.g.,][for a review]{sharma17} to biology, medicine, and statistics \citep[e.g.,][]{brooks_etal11}. 

The most common application of the MCMC approach is sampling an $n$-dimensional probability distribution function (pdf), $\pi(x)$, and to compute various related statistics, such as the average of a function $f(x)$ with respect to $\pi(x)$:
\begin{equation}
\langle f \rangle_\pi := \int f(x) \pi(x) \dx \label{eqn::avgf}
\end{equation}
where $x\in\R^n$. 
MCMC's many successes not withstanding, the computational cost of the $\pi(x)$ evaluation, slow convergence of the MCMC estimate, and the high dimensionality of $x$ often make evaluation of $\langle f\rangle_\pi$ a computationally challenging problem. This is particularly true when the low-probability tails of the distribution contribute significantly to the integral of eq. \ref{eqn::avgf} and when $\pi(x)$ is multi-modal, as is often the case in astrophysical applications in multi-dimensional spaces \citep[e.g.,][]{farr_etal14}.  For example, in sampling a Bayesian posterior in a statistical analysis, one may be interested in robust determination of credible regions up to high-levels (e.g., $99.99\%$) to evaluate the level of discrepancy with theoretical prediction or with another measurement.

\begin{figure}
\begin{center}
\includegraphics[width=.45\textwidth]{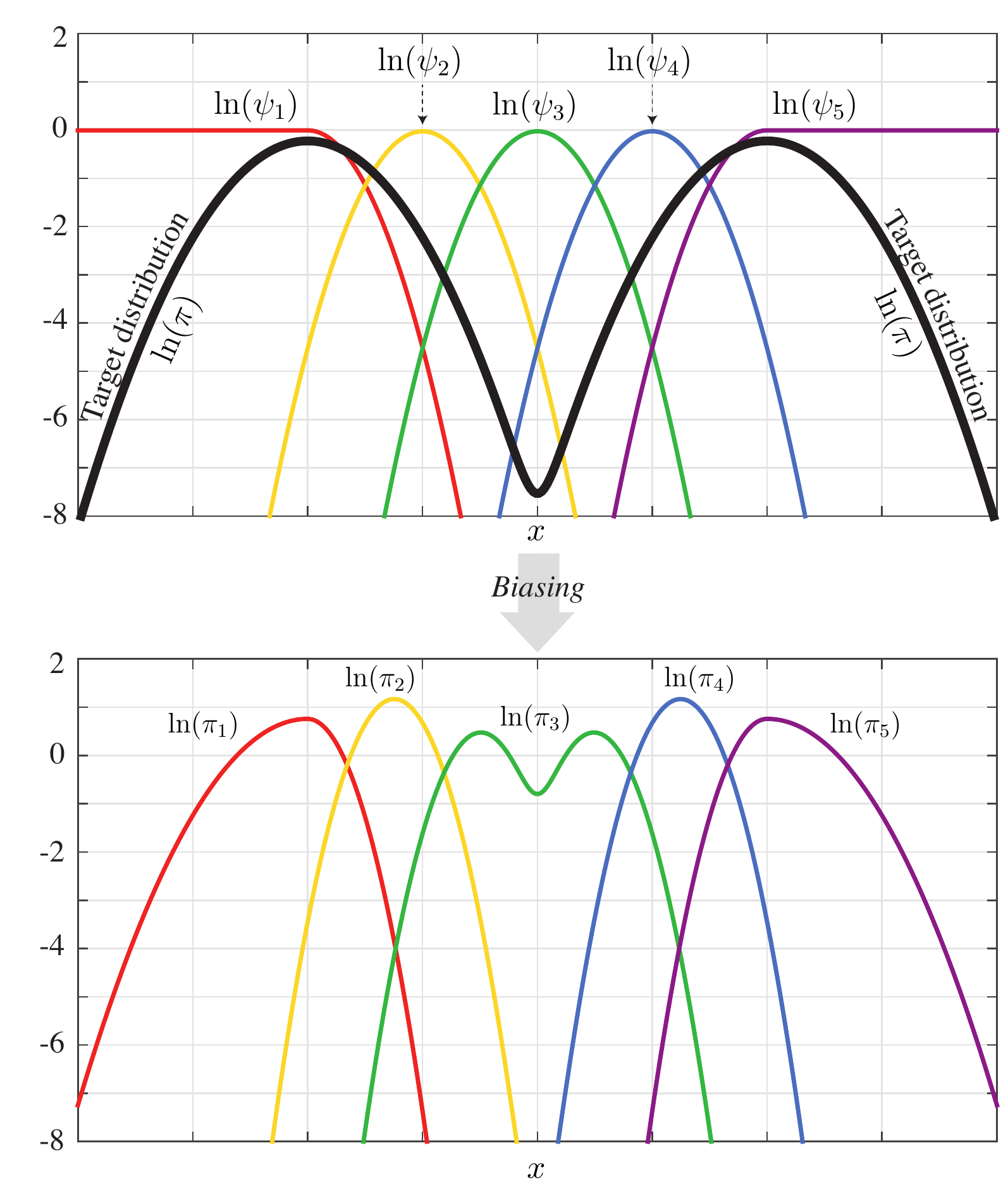}
\caption{Umbrella sampling splits the difficult problem of sampling $\pi$ (black curve) into smaller, simpler subproblems $\pi_i$ (colored curves, bottom) by introducing biasing functions $\psi_i$ (colored curves, top) so that $\pi_i \propto \pi\psi_i$. The $\pi_i$ distributions can be sampled independently, with $\pi$ recovered as a weighted sum of samples. \label{fig::usfig} 
}
\end{center}
\end{figure}

One technique in widespread use in computational chemistry  is umbrella sampling (US) -- a variant of the importance sampling approach originally proposed by \citet{torrie_valleau77}. The method has proved crucial in many chemical problems where traditional sampling methods are unable provide insight  \citep[see e.g.,][]{boczko1995first,berneche2001energetics}.
In the US algorithm, sampling of $\pi$ is split or {\it stratified} into several easier sampling problems  (see Figure \ref{fig::usfig}). Specifically, a sequence of overlapping window functions, or {\it umbrellas} $\psi_i(x)$, is introduced and the algorithm samples the corresponding  distributions, $\pi_i(x)\propto \psi_i(x)\pi(x)$. Selecting windows in low probability regions of the posterior and thereby confining samples of $\pi_i$ to these regions, allows one to get a much more efficient coverage of outlying areas of the posterior and ensure discovery of widely separated peaks in multi-modal distributions. In this sense, the umbrella sampling approach
is a part of a large class of MCMC methods that are designed to sample the parameter space more uniformly, such as parallel tempering \citep[see, e.g.,][and references therein]{earl_deem05} and parallel MCMC \citep{vanderwerken_schmidler13,basse_etal16}. 

The idea of umbrella sampling is simple enough, but its successful application relies upon a robust way to combine the samples in different umbrellas into a set of samples of $\pi(x)$   
\citep[see][and references therein]{dinner_etal17}. In particular, an efficient iterative method for computing the relative weights of the samples -- Eigenvalue Method for Umbrella Sampling (EMUS) -- has been developed \citep{thiede_etal16,dinner_etal17}.  This re-weighting is cheap and does not require extra evaluations of $\pi$. Moreover, it does not interfere with the sampling of the individual  $\pi_i(x)$ distributions. As a result, the sampling of the umbrella distributions $\pi_i(x)$ is independent of the re-weighting procedure and can be done with any MCMC sampling method that is deemed sufficiently effective for the task. 

In this paper, we present the EMUS method and its public implementation in a python package. The current version of the package uses a parallel implementation of the affine-invariant ensemble sampling algorithm of \citet[][hereafter GW10]{goodman_weare10} in the {\tt emcee} code\footnote{\href{http://dfm.io/emcee/current/}{http://dfm.io/emcee/current/}} \citep{foreman_mackey_etal13} to sample target distributions  $\pi_i(x)$. However, the US framework is general and other samplers can be used instead of {\tt emcee}. Given that sampling in different umbrellas can be done in parallel, US combined with {\tt emcee} allows us to exploit parallelism on two levels: both while sampling within individual windows and in sampling different windows independently, with occasional replica exchange communications. 

The GW10 and its implementation in {\tt emcee} are themselves quite efficient in sampling degenerate distributions and traversing relatively high-probability valleys between peaks in multi-modal distributions. Umbrella sampling is designed to make sampling of low-probability areas much easier and thus its combination with a sampler, such as {\tt emcee}, not only allows for efficient sampling of low-probability tails of distribution, but also efficient traversal of the low-probability valleys between peaks in the distributions.

This is illustrated using a distribution with two peaks in Figure \ref{fig::usfig}. In a standard scheme, samples would visit the low-probability valley between the peaks extremely rarely and thus discovery of the peaks or mixing of samples between them would be difficult. Umbrella windows placed between the peaks, on the other hand, restrict samples to these low probability valleys ensuring that many samples are available for mixing through the low-probability region. 

Clearly, the approach is most effective when we have some knowledge about the low probability regions of the target distribution, as umbrellas can be designed  specifically to sample these regions efficiently. Such information is often available either from prior knowledge or exploratory MCMC runs. 

We also  show that umbrella sampling can be efficiently applied even in  cases when no such prior information is available. In this case umbrella windows can be chosen so that the $\pi_i$ are tempered distributions, defined as in the  parallel tempering method.  US provides a mechanism by which samples from the high temperature distributions (in low probability regions) can be incorporated into more accurate estimates of tail probabilities.

The layout of the paper is as follows. In Section \ref{sec::emus} we 
describe the umbrella sampling method and the  EMUS algorithm that re-weights samples  to reconstruct the full target distribution (Section \ref{sec::weights}). Sections \ref{sec::cvbias} and \ref{sec::tempbias} detail two specific options for the windows $\psi_i$ for generic applications of umbrella sampling. Section \ref{sec::evidence} explains how we can use umbrella sampling for evidence calculation, while Section \ref{sec::parallel} describes how the algorithm is parallelized. Section \ref{sec::examples} presents two numerical examples demonstrating the utility of the umbrella sampling approach. We finish with discussion and conclusions in Section \ref{sec::conclusions}.

\section{Umbrella sampling method} \label{sec::emus}

In the umbrella sampling method the target pdf $\pi(x)$ is constructed from sampling $L$ individual distributions $\pi_i(x)$, sampled independently where
\begin{equation}
\pi_i(x) := \frac{1}{z_i} \psi_i(x)\pi(x), 
\label{eq:umbf}
\end{equation}
with umbrella window functions $\psi_i(x)$, often called the biasing functions or ``umbrellas'',  and
 normalization constants $z_i$ are defined by the condition
\begin{equation*}
z_i := \int \psi_i(x)\pi(x)\dx
\end{equation*}
ensuring that  $\pi_i$ is a properly normalized pdf ($\int\pi_i \dx=1$), even if the normalization of $\pi(x)$ is unknown. 

The calculation of these normalization constants, $z_i$, is a key part of the algorithm. Before we discuss their practical calculation, we will discuss possible strategies for windows $\psi_i$. The optimal strategy is problem specific, but 
some general strategies will nevertheless give a robust improvement in most cases. 

The simplest case is when we know a variable $x_k$ in the vector $x$ along which it is important to sample the low probability regions of $\pi(x)$.
The windows can then be defined along the $x_k$ direction: $\psi_i(x)=\psi_i(x_k)$. Similarly we can define a windowing along two or more directions: $\psi_{i,j}(x)=\psi_i(x_k)\psi_j(x_l)$. However, 
such a splitting is akin to gridding and the curse of dimensionality usually makes it impractical for more than three dimensions. 

In practice, many posteriors are difficult to sample along directions that correspond to a function of one or more $x$ components: $\sigma(x_k,\ldots)$. We shall refer to such a function as a collective variable (CV).
For example, if the posterior contains a degeneracy ridge, the difficult-to-sample direction 
is often perpendicular to the ridge, and the projection onto this direction may be a useful choice for CV with $\psi_i = \psi_i(\sigma)$. 

In the case where no intuition exists for a choice of CV, it is possible to construct an efficient method by considering $\pi_i(x)$ to be tempered distributions (see Section \ref{sec::tempbias}), similar  to those used in the parallel tempering approach \citep[e.g.,][and references therein]{earl_deem05}.

In the following subsections we describe in detail the method to estimate relative normalization constants, or {\emph {weights} $z_i$, based upon previous work by \citet[][]{vardi1985empirical,meng1996simulating,shirts2008statistically}. We also give specific illustrations  where the collective variable and $\log\pi$ stratification strategies are employed
to sample distributions. A more rigorous exposition of the approach, including error analysis and proofs of its consistency, are presented in \citet{thiede_etal16} and \citet{dinner_etal17}. 

\subsection{Calculation of window weights}
\label{sec::weights}
We seek to evaluate $\langle f \rangle_\pi$: the average of a function $f(x)$ over some pdf $\pi(x)$ defined in Equation \ref{eqn::avgf}. The goal of umbrella sampling is to recast this equation as the weighted sum over the $L$ umbrella distributions.  For a set of $L$ umbrella windows, $\{\psi_i(x)\}$,  which combined cover the entire region of space $x$ relevant for evaluation of the target integral, we can construct MCMC samples from the biased distributions $\pi_i(x)=\psi_i(x)\pi(x)$, with normalizations 
\begin{equation}
z_i = \int \psi_i(x)\pi(x)\, dx = \langle \psi_i\rangle_\pi.
\label{eq:zi}
\end{equation}

From this definition of $z_i$, we can rewrite Equation \ref{eqn::avgf} in a different form: 
\begin{eqnarray}
\langle f \rangle_\pi  &=&  \int f(x) \pi(x) \, \dx\nonumber\\ 
 &=&  \int f(x)\, \frac{\sum_{i=1}^{L} \psi_i(x) / z_i }{\sum_{j=1}^L \psi_j(x) / z_j}\,\pi(x) \, \dx\nonumber\\
 &=& \sum\limits_{i=1}^L \int\,\frac{f(x)}{\sum_{j=1}^L \psi_j(x) / z_j}\,\frac{\psi_i(x)}{z_i}\,\pi(x) \,\dx \nonumber\\
&=& \sum\limits_{i=1}^L \int\,\frac{f(x)}{\sum_{j=1}^L \psi_j(x) / z_j}\,\pi_i(x) \,\dx\nonumber\\
&=& \sum\limits_{i=1}^L\left\langle \frac{f(x)}{\sum_{j=1}^L \psi_j(x) / z_j} \right\rangle_{\pi_i}, \label{eq:us}
\end{eqnarray}
Here $\langle\rangle_\pi$ and $\langle\rangle_{\pi_i}$ denote averages with respect to distributions $\pi$ and $\pi_i$, respectively. 
Thus the average of $f$ over $\pi$, can be computed as the sum of averages of $f(x)/[\sum_{j=1}^{L} \psi_j(x) / z_j]$ over the  windows $\pi_i$.

The sum can be computed by sampling the $\pi_i$ distributions to evaluate each of the terms. Though we do not need to know the  normalizations ${z_i}$ in order to sample, they will be needed in the end 
to define relative weights of each umbrella sample. These can then be used to reconstruct the target pdf $\pi(x)$.   

The key part that is left is to define the weights ${z_i}$ for a given set of windows ${\psi_i}$ and target pdf $\pi(x)$. 
To do this, let us define a matrix $[F_{ij}]$ with elements defined as follows: 
\begin{equation}
F_{ij} =  \left\langle \frac{\psi_j / z_i }{\sum_{k=1}^{L} \psi_k / z_k} \right\rangle_{\pi_i}.
\label{eq:Fij}
\end{equation}
Thus, evaluation of $F_{ij}$ involves averaging of the normalized window functions over random MCMC samples of $\pi_i$ distributions. 
It is clear that a particular entry $F_{ij}$ will be zero if there is no overlap between samples of $\pi_i$ and support of the window $\psi_j$. 
We therefore call $F$ the {\it overlap matrix.}

The product of the vector $z=[z_1, z_2,\dots,z_{L}]$ and the $j^\textrm{th}$ column of $F$ will then be 
\begin{equation}
\sum\limits_{i=1}^{L}z_i\,F_{ij} = \sum\limits_{i=1}^{L} \left\langle \frac{\psi_j }{\sum_{k=1}^{L} \psi_k / z_k} \right\rangle_{\pi_i} = \langle \psi_j\rangle_{\pi} = z_j,
\label{eq:zF}
\end{equation}
from Equation \ref{eq:us}, and using the definition of $z_i$ in Equation \ref{eq:zi}.

Considering all columns of $F$ in Equation \ref{eq:zF} gives the left eigenvalue problem
\begin{equation}
z F(z) = z,
\label{eq::zF2}
\end{equation}
the solution of which is the required vector $z$ of normalization constants. 
Existence of the solution is guaranteed by the Perron-Frobenius theorem, if matrix $F$ is
irreducible, i.e. it cannot be transformed into block upper-triangular form by row and column  permutations -- the requirement that is satisfied if umbrella windows overlap \citep[see Section 2 in][for proof]{dinner_etal17}.

The solution of Equation \ref{eq::zF2} can be obtained using a fixed point iteration in $z$, and does not require extra sampling of the distribution. As we only need to solve for the $z$ values once, this is an inexpensive additional computation compared to the sampling of the $\pi_i$ distributions.
Note again that for $F$ to be non-degenerate, there should be a sufficient fraction of windows that do overlap. 
  
Note also that we only need to obtain the $z_i$ values up to a constant multiple, as if
\[
\widehat{z_i} = \alpha z_i
\]
for some $\alpha>0$ independent of $i$, then
\[
\sum_{i=1}^L \left< \frac{f}{\sum_{j=1}^L \psi_j / \widehat{z_j}} \right>_i = \alpha \langle f \rangle,
\]
and  $\alpha$ can be evaluated by computing the above with $f(x)\equiv1$. Note that for $L=1$ this is equivalent to the importance sampling using a biasing function $\psi_1(x)$. Thus, importance sampling can be viewed as a specific case of umbrella sampling. 

\subsection{Replica exchange} \label{sec::repex}

Replica exchange is a technique often used to enhance the rate of exploration in umbrella sampling. In this method  multiple copies of the simulation are used, known as {\emph {replicas} or {\emph{walkers}, to sample distributions $\pi_i$, with periodic exchanges of walkers between windows to promote faster mixing. Every $K$ steps, we choose a random walker $w_i$ in window $i$ and walker $w_j$ in window $j$, and swap their positions with a probability that leaves the overall target distribution intact.

Specifically, if the position of walker $w_i$ is $x_i$, and similarly walker $w_j$ is $x_j$, then the probability of accepting the swap is
\begin{align*}
P(\textrm{accept swap}) &= \min\left(1, \frac{\pi_i(x_j) \pi_j(x_i) }{\pi_i(x_i) \pi_j(x_j) } \right)\\
&= \min\left(1, \frac{\psi_i(x_j) \psi_j(x_i) }{\psi_i(x_i) \psi_j(x_j) } \right)
\end{align*}
by virtue of $\pi_i\propto\pi\psi_i$. Accepting the swap is equivalent to reassigning which window the walkers belong to, and as we only need to evaluate the bias functions this process amounts to book-keeping and is of negligible computational cost.

Exchanging walkers is most likely to succeed where the overlap between the biasing distributions is greatest, so typically we consider only swapping between adjacent windows with $j=i+1$.

In the worst case replica exchange should not harm the progress of sampling the $\pi_i$, whereas for many choices of biasing function it becomes critically important. For example, in the case of multi-modal distributions with peaks separated by very low-probability areas, replica exchange can greatly increase the efficacy of the sampling.

\subsection{Collective variable stratification} \label{sec::cvbias}

When we know the direction in the parameter space in which we want to stratify the target pdf, we can define that direction as a collective variable. CV stratification allows one to make use of prior intuition of the structure of the likelihood, or to focus sampling in  one area of interest. For example, if we know that the posterior is expected to have peaks, collective variable can be defined to be along the direction connecting the peaks. In the real-world example we will discuss below in Section \ref{sec::OmOl}, we are interested in estimating the probability that the universe is decelerating given supernova type Ia observational data. Given that we know the region in the parameter space corresponding to a decelerating universe, we can define 
umbrella windows as to maximize the efficiency and accuracy of sampling this region. 

Without loss of generality, the CV direction can be define as a function $\sigma(x)\in [0,1]$. For $L$ umbrella distributions, we define an increasing sequence of centers $c_i \in [0,1]$ where
\[
0\leq c_1 < c_2 < \ldots < c_{L} \leq 1.
\]
The biasing functions are then designed to restrain points around the associated center in CV-space. A common choice is to use Gaussians as the bias functions: 
\[
\psi_i(x) = \exp\left\{-\frac{\kappa_i^2}{2}[\sigma(x) - c_i ]^2 \right\}. 
\]  
The $\kappa_i$ defines the strength of the restraint towards the target CV value. If chosen too weakly, the sampling protocol will be ineffective, but if $\kappa_i$ is chosen too strong there will be poor overlap between windows and thus a possibility of degenerate $F$ matrix (see \S \ref{sec::weights}). 

A good balance in experiments is to choose the adjacent windows to be two standard deviations away, so
\[
\kappa_i = 2 / \max( c_i-c_{i-1} , c_{i+1}-c_i ),
\]
where $c_{-1}=0$ and $c_{L+1}=1$.

An alternative is to use tent bias functions (sometimes called chapeau functions) defined as 
\[
\psi_i(x) = \left\{\begin{array}{lc} 
1-|\sigma(x) - c_i|/ l_i & \mathrm{if } \,\, |\sigma(x) - c_i| \leq l_i \\ 
0 & \mathrm{otherwise} \end{array}\right. \tag{tent}
\] 
where the parameter $l_i>0$ defines the width of the bias' support, with a reasonable choice being $l_i = 2 / \kappa_i$. This gives a sawtooth-like family of bias functions with a steeper log-bias than the harmonic umbrella close to the edges. Tent biases have compact support which may be beneficial where the log likelihood is particularly steep preventing effective stratification. However, one disadvantage is that the initialization is more difficult without knowing $\sigma^{-1}(x)$ as sample points need to be started  inside the support of the tent.

Stratification along a collective variable coordinate can give more accurate information about e.g. the height of a barrier by concentrating samples in particular regions in space. However, the hidden degrees of freedom (i.e. the space orthogonal to $\sigma$) can stymie the progress of the sampling if the CV is chosen poorly. For example, consider a planar ring-shaped likelihood distribution, with small peaks and troughs around the ring.  Stratifying in the $x$ or $y$ direction would define a multimodal  $\pi_i$, making sampling more difficult. A more sensible CV would be to use the angle $\mathrm{atan}(y/x)$, which would break the ring into small arcs.

\subsection{Temperature stratification} \label{sec::tempbias}

If an obvious collective variable choice is not readily available  we can define the biasing functions $\psi_i$ similarly to the modified posteriors in the parallel tempering approach \citep[see][for a review]{earl_deem05}. Namely, for a series of $L$ temperatures $T_i$
\[
1\leq T_1 < T_2 < \ldots < T_L
\]
the sequence of biasing window functions is defined as
\begin{equation}
\psi_i(x) = \exp[(1/T_i-1) \log\pi(x)], 
\label{eq:psi_temp}
\end{equation}
ensuring that
\begin{equation}
\pi_i(x)\propto\pi(x)\psi_i(x)=\exp\left[\frac{1}{T_i}\log\pi(x)\right]=\pi(x)^{1/T_i},
\end{equation}
as expected in the parallel tempering approach. Higher temperatures effectively flatten distribution $\pi(x)$ allowing for exploration of wider ranges of parameters. 

This can be further improved by incorporating replica exchange into umbrella sampling simulations (see Section \ref{sec::repex} for details). A balance must be struck between the range and number of temperatures that are used. Typically we follow the advice from parallel tempering literature, and choose temperatures that are spaced exponentially ($T_k=\exp(\lambda k)$) to give roughly equal exchange probabilities between windows.

Note that although the sampling procedure in this case is identical to the traditional parallel tempering simulation (with $T_1=1$), the re-weighting scheme described in \S \ref{sec::weights} allows   us to use samples from all $L$ windows rather than just the $T_1$ window, as in the standard parallel tempering approach. 
Thus, the umbrella sampling approach greatly enhances the efficiency of the parallel tempering method. This is particularly critical for applications where evaluations of the likelihood are expensive, as is often the case in cosmology. 

Note also, the US re-weighting scheme differs from the na\"ive  importance sampling re-weighting, in which samples for different temperatures are   are combined with   the weights $\pi/\pi^{1/T_i}$. The latter is highly inaccurate, as demonstrated in Section \ref{sec::smile} (see Figure \ref{fig::smileres}). 

\subsection{Estimating evidence using US MCMC samples}
\label{sec::evidence}

In Bayesian model comparisons, one needs to evaluate the evidence, or marginal likelihood -- the integral of the posterior over the entire parameter space. 
Although a number of approaches to estimating the evidence have been explored \citep[see][for a review]{friel_wyse12}, it is particularly convenient to use the MCMC samples themselves to estimate evidence. However, the MCMC samples in the standard MCMC sampling algorithms are biased to the high-probability regions by construction. For ``diffuse'' prior distributions the contribution of low-probability areas to the evidence integral can be large \citep[e.g.,][]{efstathiou08,trotta17,cousins2017}.   US sampling of
the low-probability areas can improve the accuracy of the evidence estimates in such cases. For example, estimation of the evidence using samples from multiple biased distribution have been considered by \citet{geyer94}. 

The marginal likelihood can be estimated from the MCMC samples within umbrella windows and their weights using the estimator of \citet[][see their eq. 23; see also \S 2.2 in \citealt{robert_wraith09}]{gelfand_dey94}. Namely, if $q(x)$ is a normalized pdf with dimensionality of the posterior, and ${\tilde{\pi}}(x)$ is an unnormalized posterior with normalization constant (the evidence) $Z_\pi = \int{\tilde{\pi}}(x)\dx$, so that normalized posterior is ${\pi}={\tilde{\pi}}/Z_\pi$, we can write: 
\begin{eqnarray} 
\int q(x)\,\dx &=& \int \frac{q(x)}{{\tilde{\pi}}(x)}{\tilde{\pi}}(x)\dx = Z_\pi \int \frac{q(x)}{{\tilde{\pi}}(x)}{\pi(x)}\,\dx\nonumber\\
& =&Z_\pi \left\langle \frac{q}{{\tilde{\pi}}}\right\rangle_\pi,\ \ \ \mathrm{so}\ \ \ Z_\pi =  \left\langle \frac{q}{{\tilde{\pi}}}\right\rangle^{-1}_\pi.
\end{eqnarray}
In the context of the umbrella sampling approach presented in this paper (see Equation \ref{eq:us}), the evidence $Z_\pi$ can then be estimated as 
\begin{eqnarray}
Z_\pi = \left\langle \frac{q}{{\tilde{\pi}}}\right\rangle^{-1}_\pi=\left[\sum\limits_{i=1}^{L}\left\langle \frac{q(x)/{\tilde{\pi}}(x)}{\sum_{j=1}^{L} \psi_j(x) / z_j} \right\rangle_{\!\!\pi_i\,\,}\right]^{-1}.
\end{eqnarray}
For unimodal distributions, a good choice for $q(x)$ is a multivariate Gaussian distribution with the covariance matrix similar to that of $\pi$. In the case of multi-modal distributions one can adopt a Gaussian mixture approximation to $\pi(x)$ as a suitable $q(x)$. 
The accuracy of this evidence estimator was discussed in \citet[][see their section 2]{robert_wraith09}, who demonstrated that it is competitive in accuracy with other estimators. 

Alternatively, in the case of the temperature stratification (Section \ref{sec::tempbias}), one can use umbrella sampling and the thermodynamic integration method to estimate the evidence, as in the parallel tempering approach \citep{gelman_meng98,neal00}. 
\subsection{Parallelization} \label{sec::parallel}

Given that the sampling in each umbrella window is independent (up to periodic replica exchanges which require communication), we can exploit the independence to run in parallel on distributed systems.  In principle, the sampling within an umbrella can be achieved using any suitable sampler. 
Many modern sampling methods, such the GW10 method implemented in the {\tt emcee} code, sample multiple chains (walkers) in parallel. Thus, with umbrella sampling the parallelism both within umbrella windows and between them can be exploited. 

In the umbrella sampling python package we use in this article, parallel execution and communications are organized using a message passing interface library, as implemented in the {\tt mpi4py} python package. The 
available $C$ computing cores are split into $L$ MPI communicator groups corresponding to each umbrella window with $\lfloor L/C\rfloor$ cores in each group. The umbrella windows are sampled simultaneously in parallel, with a parallel sampler within each window making use of the $\lfloor L/C\rfloor$ cores  within the window MPI communicator.

The method itself is able to utilize all cores efficiently, with no additional computational costs when sampling, compared to conventional parallel samplers. The extra work involved in the umbrella sampling method is at the end of the simulations and requires no additional likelihood evaluations.

\section{Examples of umbrella sampling}
\label{sec::examples}
We illustrate the umbrella sampling method presented above using two example problems: 1) sampling of a difficult to sample, multi-modal synthetic pdf -- the Rosenbrock pdf with two distant Gaussian peaks (Section \ref{sec::smile}) and 2) sampling of the real world posterior distribution of the mean matter and vacuum energy density resulting from the existing constraints of the supernovae type Ia measurements (Section \ref{sec::OmOl}). The umbrella sampling code used in the first example can be found in the {\tt usample} python package.\footnote{Available at \href{https://github.com/c-matthews/usample}{https://github.com/c-matthews/usample}} The second test was run using {\tt usample} 
within the CosmoSIS cosmological analysis package \citep{zuntz_etal15}.\footnote{The umbrella sampler will be included in the next CosmoSIS version release.}

\subsection{Sampling the ``smiley'' pdf} \label{sec::smile}
\begin{figure}
\begin{center}
\includegraphics[width=.45\textwidth]{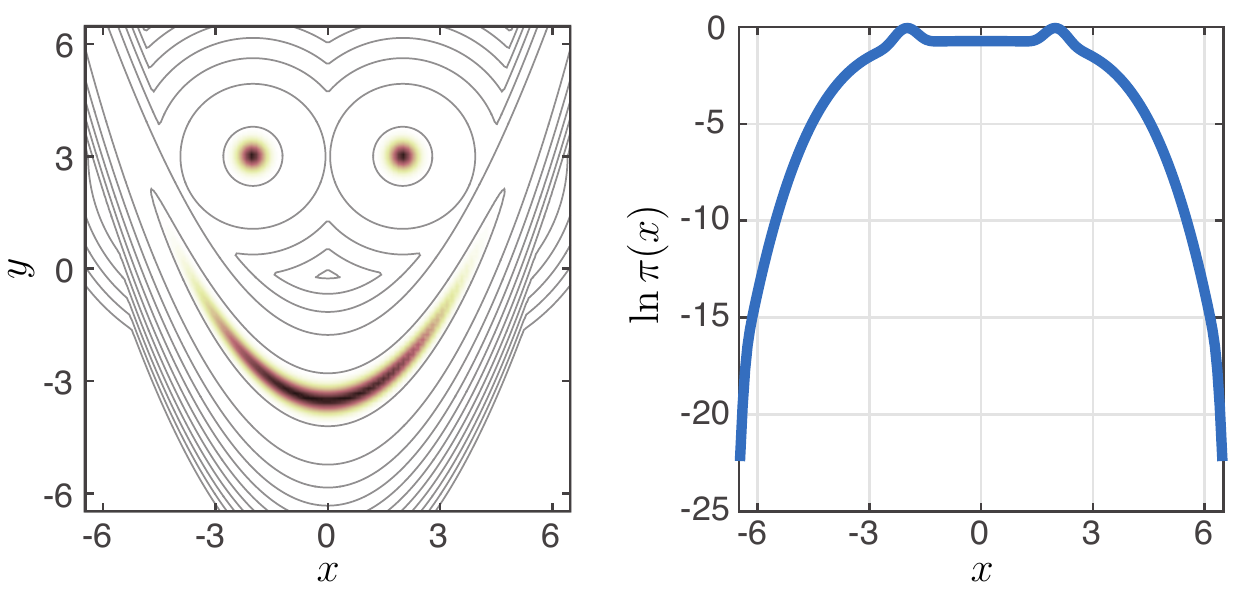}
\caption{ (\emph{Left}) The marginal posterior density is plotted in $x$ and $y$. Contours are drawn for the logarithm of the density, with the first contour at $-5$ and spaced every 25 units. (\emph{Right}) The marginal distribution is plotted in the $x$ direction.
\label{fig::smilepdf} }
\end{center}
\end{figure}
We compare results for a four dimensional toy problem, with two variables of interest (denoted $x$ and $y$) distributed in a smiley face shape, and two nuisance variables (denoted $u_1$ and $u_2$) that are distributed with independent Gaussian pdfs. The overall posterior density is
\begin{equation}
\pi(x,y,u_1,u_2) \propto \pi_\textrm{smile}(x,y) \times \exp(-u_1^2/2 - u_2^2/2),
\label{eq::smilepdf1}
\end{equation}
where $\pi_\textrm{smile}$ is
\begin{align}
\pi_\textrm{smile}(x,y) \propto & \exp(-8(x-2)^2-8(y-3)^2)\nonumber\\
&+ \exp(-8(x+2)^2-8(y-3)^2) \label{eq::smilepdf2}\\
&+ \exp(-10( y+3.5 - x^2/4)^2 - x^4/100 )\nonumber
\end{align}

The addition of the isolated distant Gaussian peaks (the eyes) to the Rosenbrock density makes the distribution multi-modal, with standard MCMC methods requiring a large number of samples to fully converge.
For example, \citet{goodman_weare10} show convergence on the 2d Rosenbrock pdf alone requires of order  $10^9$ MCMC samples even with their affine-invariant algorithm that has a relatively short autocorrelation length.

To test the efficiency of sampling for such a pdf, we will compare the performance of different sampling methods when producing an accurate log-marginal curve in the $x$ direction, plotted in Figure \ref{fig::smilepdf}. However, even though we have  good prior knowledge of the properties of $\pi_\textrm{smile}$ in this case, it is still difficult to choose an optimal collective variable $\sigma$ for this pdf along which the sampling could be stratified effectively. For example, the choices of $\sigma(x,y)=x$, $\sigma(x,y)=y$ or any simple combination of $x$ and $y$ does not remove multi-modality from the sampling in individual windows. 

A good choice for this problem is thus to use umbrella sampling with stratification in the temperature, as discussed in Section \ref{sec::tempbias}. We show results for 
the four windows with the temperature schedule of $T_i\in[1,10,100,1000]$ and replica exchange between windows every 100 steps. With this choice the replica exchange probabilities during the run were $\approx 15\%$. We use the {\tt emcee} package implementation of the \citet{goodman_weare10} algorithm with 16 walkers to sample the pdf within each window.
For performance comparison purposes we have run the algorithm for a fixed 400,000 steps for each walker. 

It is most natural to compare results of umbrella sampling in this test 
to the sampling using the parallel tempering (PT) approach with the same parameters.
In PT only samples from $T=1$ are used in the average's estimate, but  replica exchange through the higher temperatures does allow exploration of the parameter space and discovery of isolated peaks in the pdf better than the simple MCMC sampling. In this context US uses the same trajectory data from a PT run, but offers a new way of combining the data from all temperatures to give a more accurate estimate. It does not speed up sampling of the $\pi_i$ tempered distributions, but offers a more efficient post-processing of the data compared to PT.

We also compare to the case when samples from all parallel temperature samples are used, but the results are combined using a na\"{\i}ve re-weighting of samples with weights $\pi(x)/\pi^{1/T_i}(x)$, where $x$ are samples from $\pi^{1/T_i}(x)$, rather than the US weighting scheme described in Section \ref{sec::weights}. This na\"ive weighting is analogous to the weighting often used in the importance sampling approach.  Additionally, we compare against four independent runs of the {\tt emcee} sampler without the parallel tempering with the total number of samples equal to the parallel tempering runs.  All of the runs we compare thus have the same number of likelihood evaluations and hence comparable computational cost.

We compute the absolute difference in marginalized posterior $\pi(x)$ at a given value of $x$, computed by binning the samples in the interval $[0,6.5]$ into 150 equal-sized bins. For each scheme, we obtain the final posterior over ten independent runs and we use these runs to estimate the error of the average $\pi(x)$. The 2D histogram in Figure \ref{fig::smileres} shows the $\log_{10}$ of the absolute error in the log-marginal posterior $\pi(x)$ relative to the true value  as a function of $x$ and the normalized   wall-clock time. For simplicity we plot the results for positive $x$ as the results are symmetric around $x=0$.

\begin{figure} 
\begin{center}
\includegraphics[width=.45\textwidth]{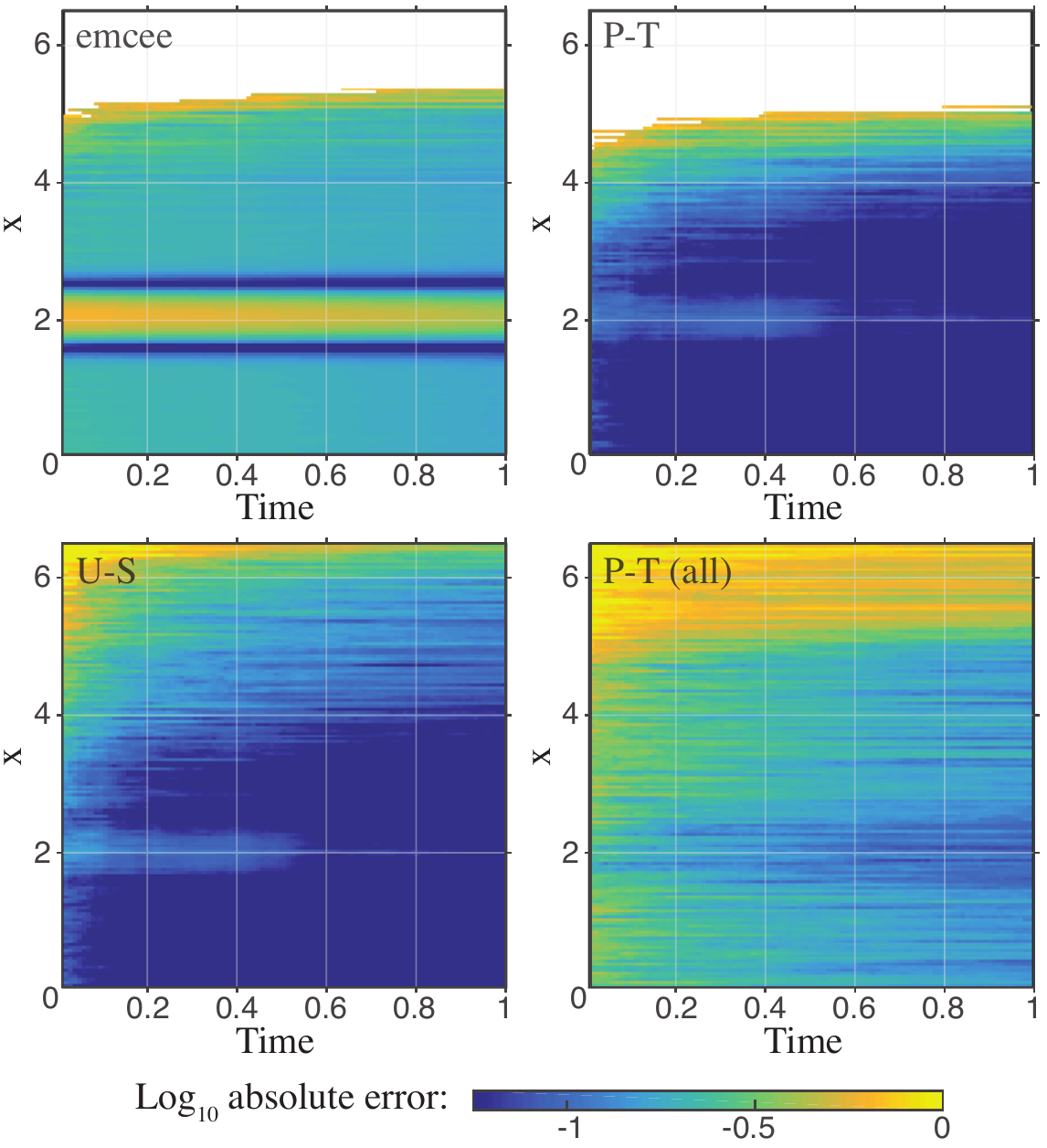}
\caption{The absolute error in the log-marginal distribution in the $x$ direction (plotted only for $x>0$), as a function of normalized wall-clock time for four sampling methods: ({\emph {clockwise}} from the top left panel) the {\tt emcee} package, parallel tempering, parallel tempering with the importance sampling re-weighting of all samples, and umbrella sampling. The color indicates the average absolute error from ten experiments, with white indicating that no samples were found at that value. 
\label{fig::smileres}}
\end{center}
\end{figure}

The GW10 scheme implemented in the {\tt emcee} package is unable to resolve the basin at $x=2$ in the allotted time, and this gives a large error that diminishes extremely slowly. This error is also apparent in the other schemes, but disappears quickly due to  replica exchange with the higher temperature simulations.

The far tails of the distribution at ${x \in [5,6.5]}$ are poorly sampled in the {\tt emcee} and PT simulations. While in regular parallel tempering we do not recover any samples in this tail, re-weighting samples from all the temperatures does give some information in this region. However, this process greatly increases the variance of the result in the entire range of $x$ and this variance does not decrease with time.

The umbrella sampling result gives the most efficient and accurate result for the entire range of $x$, even in the tail regions.

\subsection{A real-world example: cosmological constraints using type Ia supernovae}
\label{sec::OmOl}

To illustrate the power of the umbrella sampling algorithm to accurately sample the tails of the marginal posterior distribution, we use cosmological constraints derived from type Ia supernovae observations. Specifically, we sample the marginal posterior of the mean dimensionless matter and vacuum energy densities, $\Omega_{\rm m}$ and $\Omega_\Lambda$, using the SDSS-II/SNLS3 Joint Light-curve Analysis (JLA) supernovae dataset \citep{betoule_etal14} and the associated {\tt JLA v3} likelihood. 

Type Ia supernovae are one of the key probes of the cosmological parameters governing expansion of the universe \citep[see, e.g.,][for reviews]{frieman_etal08,freedman_madore10,goobar_leibundgut11} and played the main role in the discovery of the accelerating expansion of the universe \citep{riess_etal98,perlmutter_etal99}. The current supernovae samples, such as the JLA dataset, cover a wide range of redshifts and provide complimentary constraints to those derived from the Cosmic Microwave Background and the Baryonic Acoustic Oscillations measurements \citep[e.g.,][]{eisenstein_etal99}. 

Recently, the significance of the evidence for acceleration from supernovae observations alone was questioned   \citep{nielsen_etal16}. Given that estimate of such significance at the $\gtrsim 3-5\sigma$ level requires reliable sampling of the posterior tails, this problem 
is a good target for application of the US algorithm. In this section we show that the US approach allows us to make these estimates accurately and efficiently, even in the regions of the parameter space that contain a tiny fraction of the total integrated probability. 

We test two different parameterizations of windows in the umbrella sampling approach: temperature stratification only and a combination of the temperature stratification with a collective variable. The latter is possible in this case, because we are interested in estimating the probability
that the universe is not accelerating, and  we can
define the collective variable in the direction perpendicular to the
line separating accelerating and non-accelerating universes: $\Omega_{\rm m}/2 - \Omega_\Lambda=0$. 

When using only temperature windows, we use the following schedule of sixteen temperatures: $T_j=50^{(j-1)/15}$. We believe that this is a robust and flexible choice when little about the overall likelihood surface is known a priori, as it will give thorough sampling in all variables.

When using the collective variable we also use sixteen windows, but each window uses one of four temperatures (with $T_i\in\{1,3.7,13.6,50\}$) as well as one of four collective variable centers (with $c_j\in\{0,1/3,2/3,1\}$), divided so each window has a unique pair $(T_i,c_j)$. The overall bias function for a window is then the product of the bias functions in the collective variable and in the temperature. We use a tent bias function with the collective variable
\[
\sigma(x) = \max\{0,\min[1 , (x-p_1) \cdot (p_2-p_1) / \|p_2-p_1\|^2 ]\},
\]
which gives a normalized distance of a point along a line connecting two anchor points $p_1$ and $p_2$, when the point $x$ is projected along the line. The anchor points are chosen so that the line segment they define is perpendicular to and intersecting the line separating accelerating and decelerating universes, $\Omega_{\rm m}/2 - \Omega_\Lambda=0$: 
\[
p_1 = (0.55,0.9,0,\ldots)^T,\quad p_2 = (0.85,0.3,0,\ldots)^T. 
\]
This ensures that the level sets of $\sigma(x)$ define strips parallel to the line $\Omega_{\rm m}/2 - \Omega_\Lambda=0$. 

\begin{figure}
\begin{center}
\includegraphics[width=.4\textwidth]{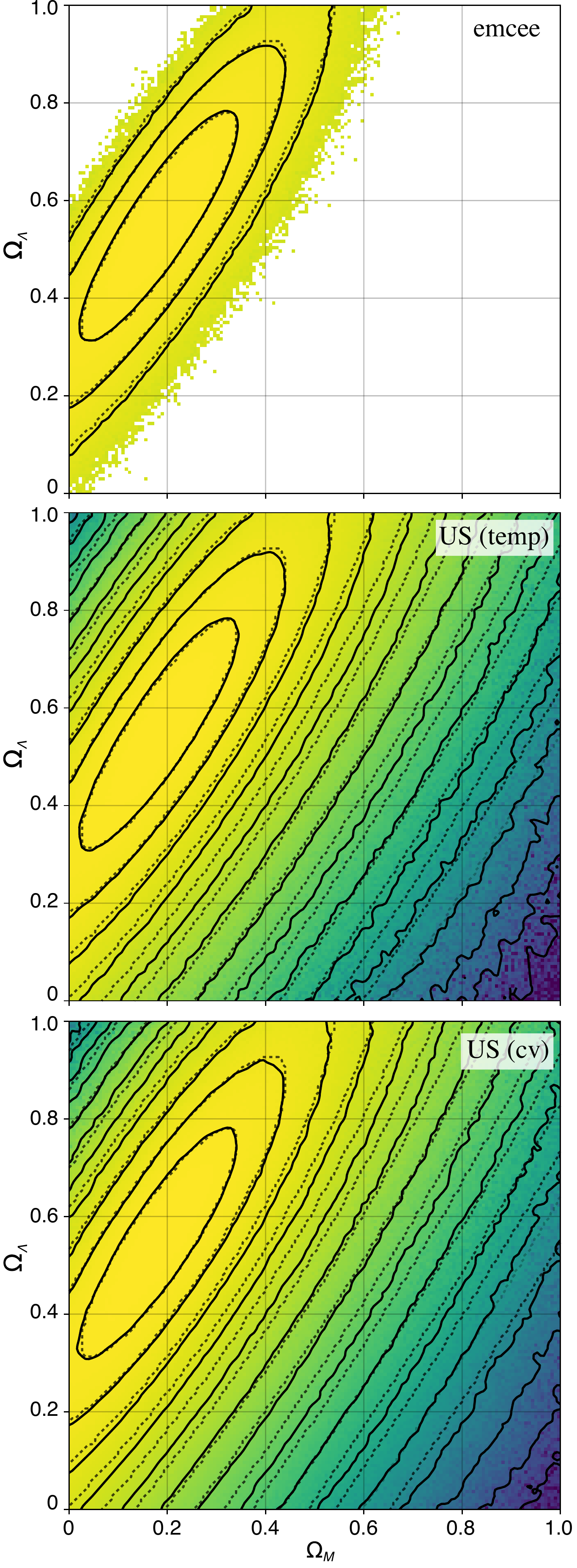}
\caption{The iso-density contours of the sampled marginal posterior distribution using the {\tt JLA v3} likelihood in the $(\Omega_m,\Omega_\lambda)$ plane, marginalized over the 6-dimensional space in which sampling was done.The three panels show results of sampling using the \protect\citet{goodman_weare10} algorithm implemented in the {\tt emcee} code (top panel) and umbrella sampling with different choices of the umbrella partition (middle and bottom panels). In each case, the sampled was done using similar amount of CPU time.   The solid lines show the actual iso-significance contours of the posterior up to fifteen sigma levels, while the dotted lines give the corresponding contours for a Gaussian approximation of the posterior. A Gaussian filter was used to smooth the plotted solid contours, but no filter was applied to the underlying shaded surface. \label{fig::3panel}} 
\end{center}
\end{figure}

Using this collective variable $\sigma(x)$ means that some samples will be drawn from within the target region in the tail of the distribution. This improves
efficiency and accuracy of the estimate of the probability that we are interested in:
\begin{equation}
p_{\rm dec} = \int \mathbf{1}_\mathrm{dec}(x) \pi(x)\dx,
\label{eq::pdec1}
\end{equation}
where $x=\{\Omega_{\rm m},\Omega_\Lambda\}$, $\pi(x)$ is 
the posterior marginalized over all other parameters, and 
\begin{equation}
\mathbf{1}_\mathrm{dec}(x) = \begin{cases}
  1, & \mathrm{if } \, \, \Omega_{\rm m} > 2\,\Omega_\Lambda \\ 
  0  & \mathrm{otherwise.} 
\end{cases}.
\label{eq::pdec2}
\end{equation}

By contrast, using pure temperature windows places no such constraint on the sample distribution. 
In this case samples in each window will explore the entirety of the parameter space.

Figure \ref{fig::3panel} shows results of sampling the {\tt JLA v3} likelihood in 6D parameter space using the \citet{goodman_weare10} algorithm implemented in the {\tt emcee} code \citep[][top panel]{foreman_mackey_etal13} and umbrella sampling with different choices of the umbrella partition (middle and bottom panels). Here we sample the {\tt JLA v3} likelihood using the CosmoSIS package \citep{zuntz_etal15}, to which we have added the US sampler. In each case, the sampling was done using the same number of likelihood evaluations, and thus similar amount of CPU time. 

Calculations were carried out using two Intel E5-2670 16-core nodes with MPI communications between and within nodes using MPI pools, as described in Section \ref{sec::parallel}. For the {\tt emcee} package, we used 192 walkers with $10^5$ steps per walker. For the umbrella sampling, we used 16 windows with 32 walkers per window and $3.75\times10^4$ steps per walker. Thus, in both experiments $1.92\times10^7$ samples were 
generated.  

The runs using {\tt emcee} used all 32 available cores in parallel. The  umbrella sampling runs  used sixteen windows sampled in parallel, each with two cores that worked in parallel. This gave the umbrella sampling results the same parallel efficiency as the {\tt emcee} sampler. An mpi pool object was used to na{\"i}vely distribute the tasks evenly, without any load balancing. 

Figure \ref{fig::3panel} shows the successive sigma level contours (solid black lines). In the middle and bottom panels that show results of US method, the contours are shown to $15\sigma$. A Gaussian filter was used to smooth the plotted contours, but no filter was applied to the underlying shaded surface showing the posterior distribution.The figure shows that with a given CPU time GW10 algorithm samples the likelihood well only to $\approx 3\sigma$ contour, while the US sampler samples it with a nearly uniform accuracy to $\approx 15\sigma$ level. 

The results obtained using umbrella sampling with only temperature show a higher variance than results using collective variables, as is evident from the spikier and less-defined contours at the furthest sigma levels. This is simply because umbrella windows with the collective variable stratified, 
ensure that a fixed number of walkers sample the low probability areas of the posterior in the $\Omega_{\rm m}-\Omega_\Lambda$ plane, while in the temperature windows walkers explore the entirety of the parameter space without any restraints.  

However, the contours do demonstrate good agreement even in the tails. In particular, the first three contours show good agreement between all of the methods. This means that even US sampling with the more flexible temperature umbrellas is as accurate as the affine-invariant MCMC for the $\approx 3\sigma$ credible region, but is far more accurate in lower 
significance regions. Thus, the accuracy gain in these low probability regions is obtained without significant loss of accuracy
in high probability region.

In Table \ref{tab::area_estimate} we compare estimates for $p_{\rm dec}$ computed using sampling and the estimate using the Gaussian approximation of the posterior. We can see that the value of $p_{\rm dec}$ estimated using the {\tt emcee} sampling is in good agreement with the estimates obtained using the US method because we have allowed for sufficient length of the chains to sample the $\approx 3\sigma$ region. Nevertheless, the US sampling using collective variable achieves a factor of four smaller error for the same amount of work. Given that Monte Carlo estimates
of such quantities carry $\mathcal{O}(N^{-1/2})$ error for $N$ samples, this means that {\tt emcee} would need to 
be run with 16 times more samples to achieve the same accuracy in this case. 

\begin{center}
\begin{table*}
  \begin{tabular}{|l|c|c|}
    \hline
    Scheme & $p_{\rm dec}$ for $\Omega_{\rm m}>0$& $p_{\rm dec}$ for $\Omega_{\rm m}>0.2$ \\ \hline\hline
    Gaussian approximation & $7.3\times10^{-7}$ & $1.6\times10^{-10}$ \\  
    {\tt emcee} & $3.1\times10^{-6}\pm1.3\times10^{-7}$ & not available \\
    Umbrella sampling (temperature) & $3.3\times10^{-6}\pm1.5\times10^{-7}$ & $5.4\times10^{-9}\pm1.7\times10^{-10}$  \\  
    Umbrella sampling (CV) & $3.2\times10^{-6}\pm3.0\times10^{-8}$ & $5.4\times10^{-9}\pm4.1\times10^{-11}$ \\  
    \hline
  \end{tabular}\caption{The estimated value of $p_{\rm dec}$ is computed from the mean of five runs of each scheme, with the standard error of the estimate. \label{tab::area_estimate} }
  \end{table*}
\end{center}
The faint dotted lines in Figure \ref{fig::3panel} correspond to the contours assuming a Gaussian approximation of the posterior. More precisely, we measured the covariance parameters using the {\tt emcee} samples taken within the second sigma contour. The sampled log likelihood surface was fit to a quadratic form using the Matlab \emph{fit} function, and its corresponding sigma levels plotted. Although the initial agreement up to the third contour is good, it is clear that the Gaussian approximation fails to accurately describe the tails of the posterior. The sampled surface obtained from umbrella sampling shows that there exists a much ``fatter'' tail compared to the Gaussian.

As a consequence of this, any  estimate of the probability of a decelerating universe, $p_{\rm dec}$ (eqs \ref{eq::pdec1}-\ref{eq::pdec2}), 
 using the Gaussian posterior assumption, as  done by \citet{nielsen_etal16} for example, will be inaccurate. 
Indeed, the Gaussian approximation estimate shown in Table \ref{tab::area_estimate}  underestimates the probability that we live in a decelerating universe by a factor of six. To be precise, \citet{nielsen_etal16} estimated $p_{\rm dec}$ using the $\chi^2$ approximation ($p_{\rm cov}$ in their Equation 10), which effectively assumes that posterior is Gaussian, and using the likelihood profile rather than marginalized posterior distribution. This is not equivalent to integrating the probabilities using marginalized Gaussian posterior in the $\Omega_{\rm m}-\Omega_\Lambda$ plane. 


\begin{figure} 
\begin{center}
\includegraphics[width=.415\textwidth]{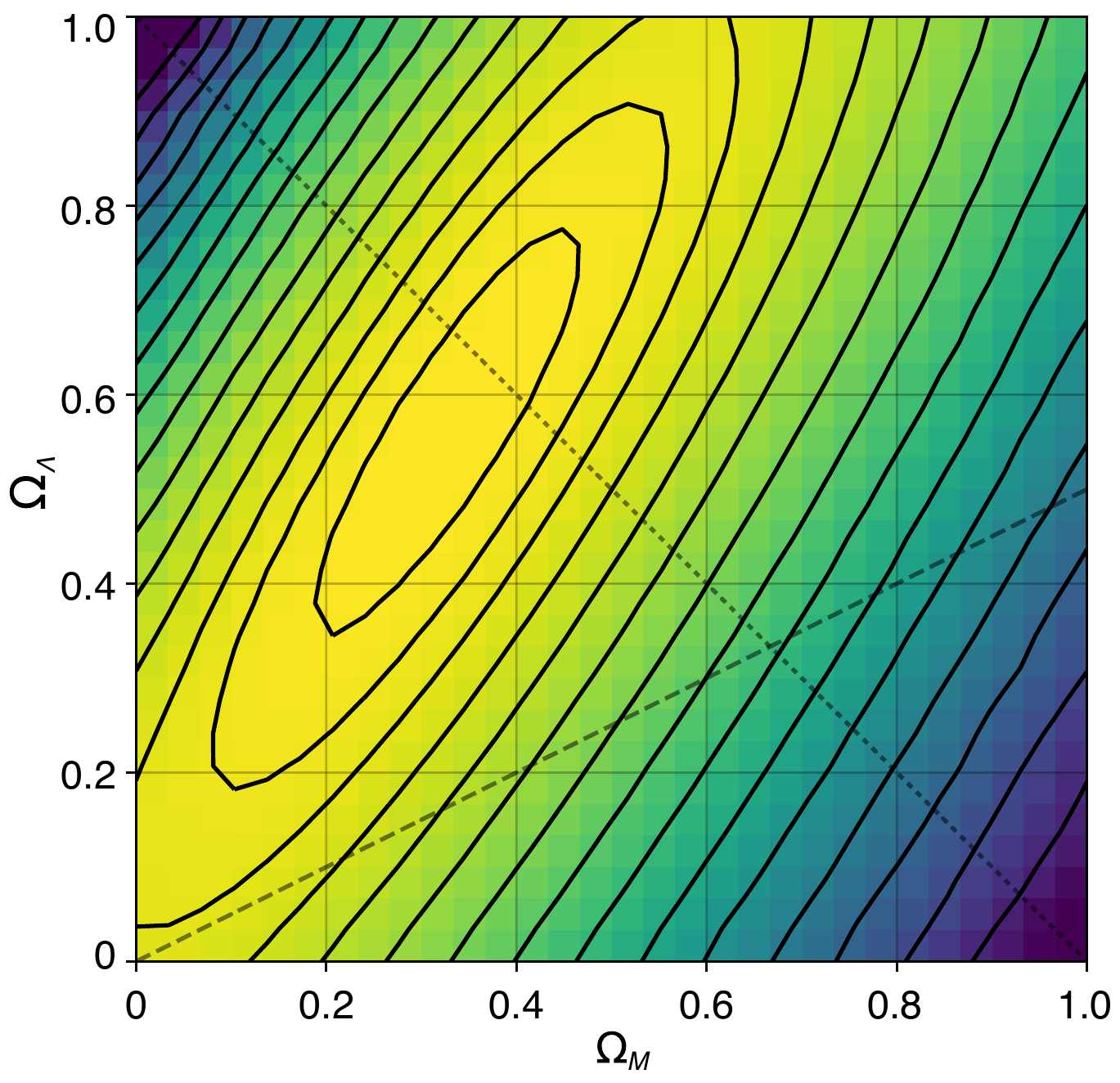}
\caption{Marginalized posterior in the $\Omega_{\rm m}-\Omega_\Lambda$ plane using the same JLA data set but now with the likelihood of \citet{nielsen_etal16}. Here we sampled the posterior varying only three parameters $M_0$, $\Omega_{\rm m}$, and $\Omega_\Lambda$, while keeping the other parameters of their likelihood fixed at the best values reported in the first row of Table 1 in \citet{nielsen_etal16}. This approximates the "profile likelihood contours" used by these authors. Note that the contours shown in this plot are the actual contours of the posterior enclosing a given fraction of the total probability, not the contours of a Gaussian pdf. The dashed line defined
by $\Omega_{\Lambda}=\Omega_{\rm m}/2$ is the boundary between accelerating (above) and decelerating (below) universes. The dotted line  is the line of geometrically flat universes, $\Omega_{\rm m}+\Omega_\Lambda=1$. \label{fig::nielsen}}
\end{center}
\end{figure}

In Figure \ref{fig::nielsen} we show the result of sampling the likelihood of \citet{nielsen_etal16}\footnote{We include the routine implementing likelihood similar to that of Nielsen et al. (2016) that 
we use in this analysis with the public version of the US code. } 
varying $M_0$, $\Omega_{\rm m}$, and $\Omega_\Lambda$, while keeping the other parameters of their likelihood fixed at the best values reported in the first row of their Table 1. This procedure approximates the likelihood profile analysis of \citet{nielsen_etal16}. Using this posterior we estimate the probability of deceleration of $p_{\rm dec}\approx 1.03\times 10^{-4}$ corresponding to $\approx 3.85\sigma$ significance, which is close to, albeit somewhat higher than, the significance reported by \citet{nielsen_etal16}. 

We can see that, compared to the JLA likelihood results in Figure \ref{fig::3panel}, the contours are shifted up and to the right towards the deceleration region and this region is now close to $\approx 3\sigma$ contour. 
This difference is the reason   values for the $p_{\rm dec}$ probability in our Table \ref{tab::area_estimate} are very small and do not show significant evidence for deceleration. The difference is due to our use of the {\tt JLA v3} likelihood, while \citet{nielsen_etal16} used a different likelihood that accounted for intrinsic scatter of supernovae properties, which is done only in post-processing and in approximate fashion in the JLA likelihood. On the other hand, \citet{nielsen_etal16} did not account for significant survey selection effects in their analysis that affect the apparent properties of supernovae. These differences likely account for the discrepancy in the
estimates of the deceleration probability. 

Regardless of these differences, other cosmological probes indicate that $\Omega_{\rm m}>0.2$ with very high confidence \citep[e.g.,][]{planck16,alam_etal17}. The probability of deceleration with $\Omega_{\rm m}>0.2$ prior for the posterior derived from \citet{nielsen_etal16} likelihood (Fig. \ref{fig::nielsen}) is only $p_{\rm dec}\approx 1.9\times 10^{-6}$ or $\approx 4.75\sigma$.   
For the JLA likelihood $p_{\rm dec}$   is only $\approx 5.4\times 10^{-9}$. Note that we cannot make this estimate with the {\tt emcee} run of the same length because it has no samples in the region of decelerating universes bounded by $\Omega_{\rm m}>0.2$. The US   runs, on the other hand, still give a reasonably accurate estimate, as shown in Table \ref{tab::area_estimate}. 

The value of the $p_{\rm dec}$ estimate as a function of the number of likelihood evaluations (equivalently simulation time) is shown in Figure \ref{fig::area_estimate}. The solid lines indicate the mean of the five independent runs, while the shaded regions indicate the standard deviation of the runs themselves, showing the expected behavior of one trajectory. 
It is clear that umbrella sampling using the collective variable provides a rapid and precise estimate for $p_{\rm dec}$. By contrast, the emcee estimate has extremely large variance for a significant portion of the run, with no samples recorded for the first tenth of the run overall. Using US with the temperature gives behavior in between these two schemes, with an early accurate approximation. However, the variance appears to decay more slowly in this case compared to the CV-defined umbrella windows.

\begin{figure} 
\begin{center}
\includegraphics[width=.47\textwidth]{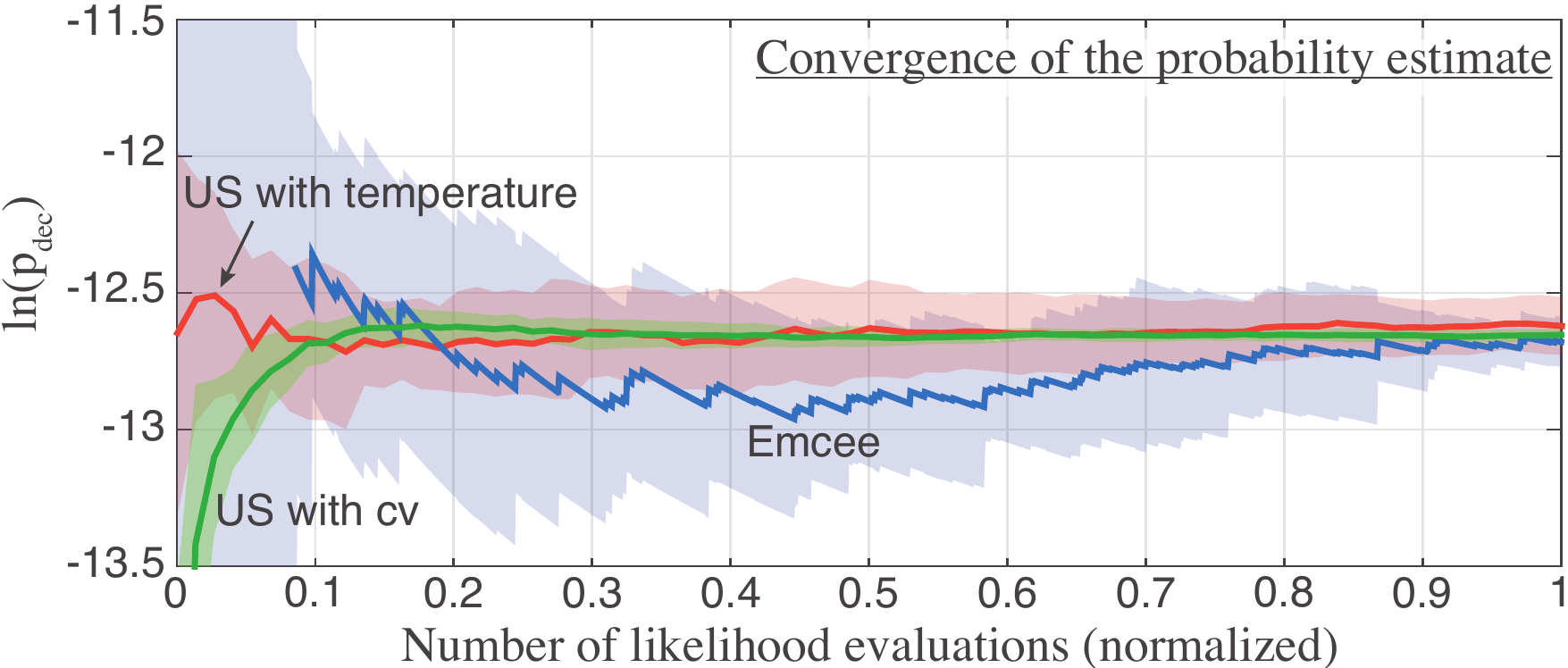}
\caption{The mean (bold line) and standard deviation (shaded region) of the estimate of $\ln(p_{\rm dec})$ is plotted as a function of the number of likelihood evaluations. Values are compared for the {\tt emcee} package (blue) versus using Umbrella Sampling with either temperature (red) or a collective variable (green). \label{fig::area_estimate}}
\end{center}
\end{figure}

\section{Discussion and conclusions}
\label{sec::conclusions}

In this paper we presented the umbrella sampling technique and showed that it can be used to sample low probability areas of the posterior distribution  that may be required in statistical analyses of data. In this technique 
the parameter space is partitioned into   umbrella windows by splitting the target likelihood into separate likelihoods  given by the original likelihood multiplied by  appropriately
weighted window functions. 
Though US has been used successfully in computational chemistry, it has not been used there to compute general averages, tail probabilities of $\pi$, or for parameter estimation. The tempering umbrella sampling  approach is, to  our knowledge, presented here for the first time.

We show that the US method is cheap and can be easily implemented ``on top'' of existing MCMC samplers, such as 
{\it emcee}. The method allows the user to capitalize on their own intuition  by using collective variables to define umbrella windows, or to make use of a more general technique by stratifying in the temperature. A publicly available standalone python package implementing the scheme can be found at:
\href{https://github.com/c-matthews/usample}{https://github.com/c-matthews/usample}. 
Additionally we have added umbrella sampling to the CosmoSIS package \citep{zuntz_etal15}\footnote{\href{https://bitbucket.org/joezuntz/cosmosis/}{https://bitbucket.org/joezuntz/cosmosis/}} 
and the US sampler will be included in the next release of CosmoSIS, as one of the available samplers. 

We presented a number of tests illustrating the power of the US method in sampling low probability areas of the posterior. We also showed that this ability allows a considerably more robust sampling of multi-modal distributions compared to the {\tt emcee} direct sampling methods. For the toy model distribution given by the sum of Rosenbrock 
and two multi-variate isolated Gaussian pdfs, the umbrella sampling method presented in Section \ref{sec::emus} was shown to be more efficient compared to parallel tempering, as well as a naive recombination of the data, despite using exactly the same set of samples. This is because in parallel tempering only one subset of samples with $T=1$ is actually used for the final analyses. By contrast, the umbrella sampling approach allows the 
use of samples from {\it all} of the temperatures. 

In the supernova cosmological constraints example, umbrella sampling was shown to provide significantly more information about the posterior in the low probability areas compared to the direct sampling by the {\tt emcee} code. In particular, as shown in Figure \ref{fig::3panel}, for the same amount of work {\tt emcee} samples the posterior to the 
$\approx 3\sigma$ credible region, while the US method samples the same posterior to the $\approx 15\sigma$ region.  This ability to sample far into the tail of the posterior distribution may find other applications, such as
evaluation of the marginal likelihood, also known as the Bayesian evidence, which requires evaluation of the integral of the posterior over the entire parameter space, as discussed in Section \ref{sec::evidence}. 

Finally, in the era of precision cosmology, as errors of cosmological parameter estimates shrink, the need to evaluate discrepancies and significance of tensions at a $\approx 4-5\sigma$ level will become commonplace. The umbrella sampling method presented in this work will allow us to do such evaluations efficiently.

\section*{Acknowledgments}
AK is very grateful to Rick Kessler and Dan Scolnic for many enlightening discussions on the cosmological analyses using supernovae type Ia samples and effects of survey selection function. 
AK was supported by a NASA ATP grant NNH12ZDA001N, NSF grant AST-1412107, and  by the Kavli Institute for Cosmological Physics at the University of Chicago through grant PHY-1125897 and an endowment from the Kavli Foundation and its founder Fred Kavli. 
EJ is supported by the Argonne Leadership Computing Facility, a U.S. Department of Energy,Office of Science User Facility operated under Contract DE-AC02-06CH11357. 
CM and JW were supported by the U.S. Department of Energy, Office of Science, Office of Advanced Scientific Computing Research (ASCR) under Contract DE-AC02-06CH11347. 
JW was also supported by ASCR through award DE-SC0014205.

Numerical experiments presented in this paper were carried out on the {\tt midway} computing cluster maintained by the University of Chicago Research Computing Center. We acknowledge the Center and its staff for support. 


\bibliographystyle{mn}
\bibliography{us}

\label{lastpage}

\end{document}